\documentclass{optica-article}

\journal{opticajournal}

\articletype{Research Article}

\usepackage{lineno}
\usepackage{subcaption}
\usepackage{graphicx} 
\usepackage{subcaption}
\usepackage{upgreek}

\begin{document}

\title{A ultrabright, two-colour photon pair source based on thin-film lithium niobate for bridging visible and telecom wavelengths}

\author{Silia Babel,\authormark{1,2,*} Laura Bollmers,\authormark{1,2} Franz Roeder,\authormark{1,2} Werner Ridder,\authormark{1,2} Christian Golla,\authormark{2} Ronja Köthemann,\authormark{1,2} Bernhard Reineke,\authormark{2} Harald Herrmann,\authormark{1,2}  Benjamin Brecht,\authormark{1,2} Christof Eigner,\authormark{2} Laura Padberg,\authormark{1,2}  and Christine Silberhorn\authormark{1,2}}

\address{\authormark{1} Paderborn University, Integrated Quantum Optics, Warburger Str. 100, 33098 Paderborn, Germany\\
\authormark{2}Paderborn University,
Institute for Photonic Quantum Systems (PhoQS), Warburger Str. 100, 33098 Paderborn, Germany}

\email{\authormark{*}silia.babel@upb.de} 

\begin{abstract*} 
We present the design and characterisation of a guided-wave, bright and highly frequency non-degenerate parametric down-conversion source in thin-film lithium niobate. The source generates photon pairs with wavelengths of 815$\,\mathrm{nm}$ and 1550$\,\mathrm{nm}$ linking the visible wavelength regime with telecommunication wavelengths. We confirm the high quality of the generated single photons by determining a value for the heralded second-order correlation function as low as $g^{(2)}_h(0) = (6.7\pm1.1)\cdot10^{-3}$. Furthermore, we achieve a high spectral brightness of 0.44$\cdot$10$^{7}$$\frac{\text{pairs}}{\text{s} \cdot \text{mW} \cdot \text{GHz}}$ which is two orders of magnitude higher than sources based on weakly guiding waveguides. The almost perfect sinc-shape and the strong agreement between the effective and nominal bandwidth highlights the success of our integrated workflow, which begins with device design and continues through precise fabrication to detailed quantum-state characterization. This comprehensive approach enables targeted optimization of the source, resulting in excellent quantum state generation. Our results set a new standard for on-chip, non-degenerate photon-pair sources and represent a crucial step towards practical, scalable quantum communication networks and photonic quantum computing.
\end{abstract*}

\section{Introduction}

Single photons are a crucial resource for future quantum technologies, such as quantum communication, quantum computing or quantum simulation, because they constitute the information carrier for light based quantum encoding.
Parametric down-conversion (PDC) sources have become a standard tool for the generation of heralded single photons and correlated photon pairs, where the photon's properties, such as wavelength, bandwidth, etc. are highly designable and hence can easily be adjusted for each specific application \cite{caspani2017integrated}. 
Moreover, a high brightness of the photon pair source is required for measurements with a high signal to noise ratio in a reasonable measurement time.\\
An emerging high potential material for the integration of bright, high quality photon pair sources is thin-film lithium niobate (TFLN). This platform is highly promising for quantum photonic integrated circuits (QPICs) and various devices such as low-loss waveguides \cite{zhu2024twenty,luke2020wafer,shams2022reduced}, electro-optic modulators \cite{xu2022dual,zhang2022systematic,hu2025integrated} and nonlinear processes \cite{zhu2023sum,wang2023quantum} have already been demonstrated. 
TFLN consists of a thin layer of LN (between 300$\,\mathrm{nm}$ and 1000$\,\mathrm{nm}$ thick) which is bonded on an insulator layer, e.g., silicon dioxide SiO$_2$, followed by a handle substrate. The strong prospects of TFLN for integrated quantum optics is due to the combination of the outstanding optical properties of lithium niobate and the high mode confinement of nano-structured waveguides resulting from the strong index contrast between LiNbO$_3$, SiO$_2$ and air \cite{zhu2021integrated}. Thus, TFLN devices exhibit on the one hand, a wide transparency window, large electro-optical coefficients and high second-order nonlinearity due to intrinsic material properties \cite{weis1985lithium}, but on the other hand additionally benefit from the very small geometry, which further boosts the nonlinear interaction in frequency mixing processes.
For quantum light sources the high modal confinement in TFLN, hence, results in a remarkable enhancement of the brightness and compactness.\\
Different PDC sources in TFLN have already been realized \cite{xin2022spectrally, ma2023highly,harper2024highly,zhao2020high,elkus2019generation,williams2024ultra,xue2021ultrabright,zhang2023scalable, javid2021ultrabroadband}, but the sources demonstrated to date are degenerate PDC sources with identical signal and idler wavelengths. However, for many applications in integrated quantum optics a highly non-degenerate PDC source offers a unique functionality for future quantum technologies. One example is in the field of networks in quantum communication. Here, a two-colour source enables the implementation of cost-effective, highly efficient heralded single photon sources where the photon with the shorter wavelength can be used to herald the existence of the other photon by measuring it with low-cost silicon avalanche single photon detectors. Moreover, the source can also be used in the framework of quantum frequency transduction \cite{huang1992observation,tanzilli2005photonic}. There the correlated photons generated at two different colours can be used to bridge stationary quibts at visible wavelengths with flying qubits at telecom \cite{bussieres2014quantum}. In this way two-colour sources can substitute frequency converters.\\
Another example is in the research field of quantum spectroscopy. Here, such a source is ideal for quantum measurements with undetected photons \cite{mukamel2020roadmap, paterova2018measurement,kalashnikov2016infrared,lindner2021nonlinear,kaufmann2022mid} in which the shorter wavelength is used for detection while the longer wavelength is used to interact with an object under test. This allows one to measure at technically accessible wavelengths in the visible or near infrared while gaining insights into an object's properties in the infrared.\\
All these mentioned applications require that the source is bright and generates highly non-degenerate photon pairs. Thus, in this paper, we present a highly non-degenerate photon pair source in TFLN, which combines compactness and an exceptional brightness with a large  span of frequencies between the generated signal and idler photons. We will present the design and describe the fabrication of the source. Characterizing our source we  find that the generated photon pairs at $\lambda_i = 1550\,\mathrm{nm}$ for the idler photon and $\lambda_s = 815\,\mathrm{nm}$ for the signal photon can be tuned via temperature with -0.472$\,\mathrm{nm/K}$ for the signal and 1.57$\,\mathrm{nm/K}$ for the idler photons. The bandwidth of the nearly ideal sinc-like measured PDC spectrum agrees with the simulated bandwidth very closely. For a further characterization of the phase matching we investigate the reverse process and find a good agreement between measurement and simulation. When benchmarking the quantum performance of our source we observe a value of the heralded second-order correlation function $g^{(2)}_h(0) = (6.7\pm1.1)\cdot10^{-3}$ proving the source generates high quality heralded single photons. Lastly, we show a high spectral brightness of B = 0.44$\cdot10^7$$\,\frac{\text{pairs}}{\text{s} \cdot \text{mW} \cdot \text{GHz}}$.

\section{Modelling and Sample Fabrication}
\label{SimulationsandFabrication}
\begin{figure}[h]
    \centering
    \includegraphics[width=9cm]{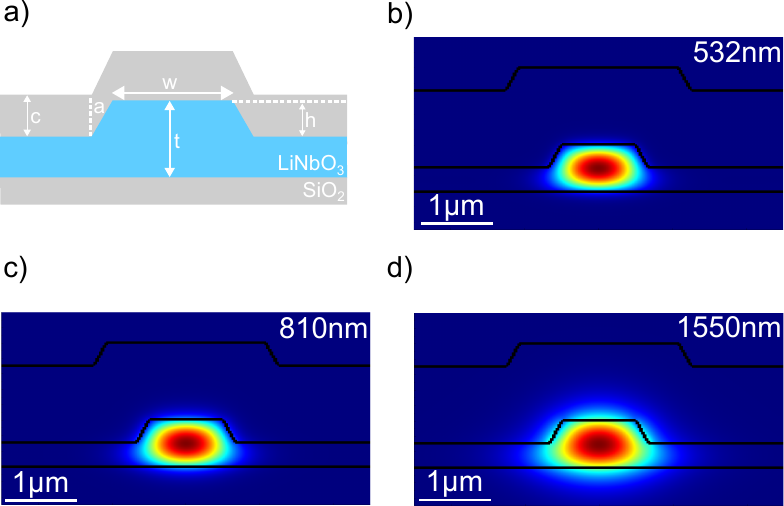}
    \caption{a) Sketch of a cross section of an etched waveguide in TFLN. b) - c) Guided TE modes of the simulated waveguide structure. The modes are simulated with the commercial software Lumerical.}
    \label{modes}
\end{figure}

\begin{figure}[h]
    \centering
    \begin{subfigure}{0.45\textwidth}
        \centering
        \includegraphics[width=\textwidth]{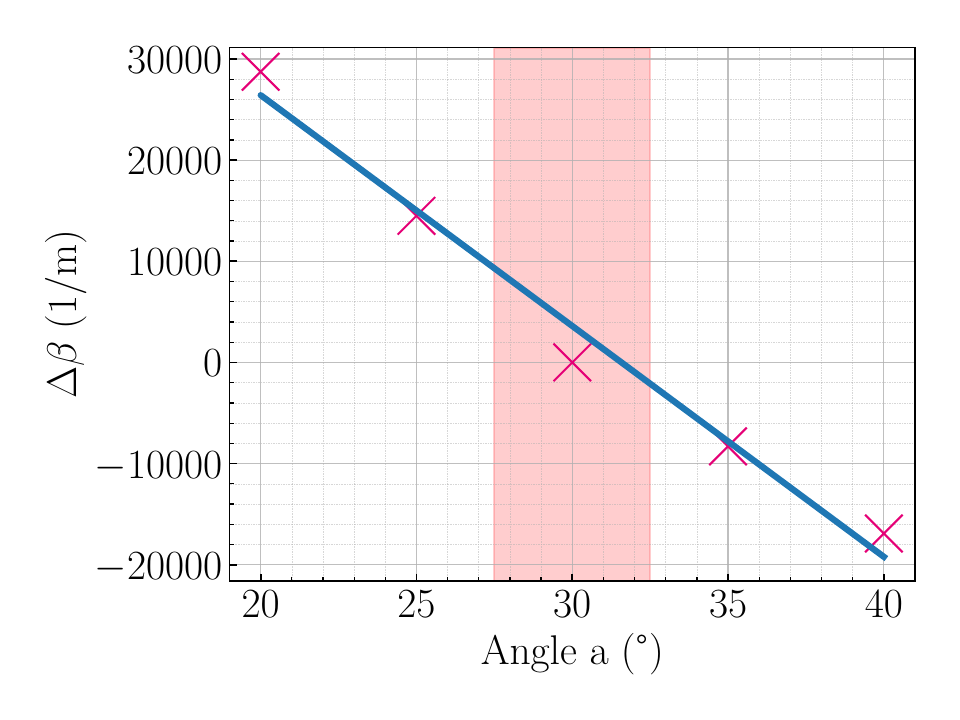}
        \label{fig:bild1}
    \end{subfigure}
    \begin{subfigure}{0.45\textwidth}
        \centering
        \includegraphics[width=\textwidth]{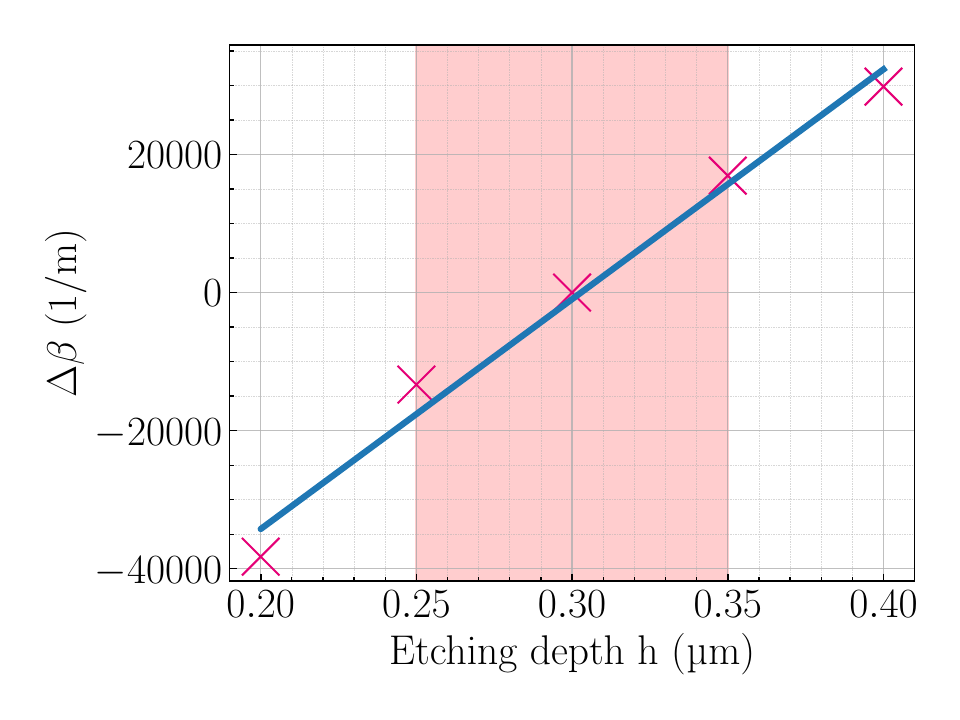}
        \label{fig:bild2}
    \end{subfigure}

    \vspace{-0.5cm}

    \begin{subfigure}{0.45\textwidth}
        \centering
        \includegraphics[width=\textwidth]{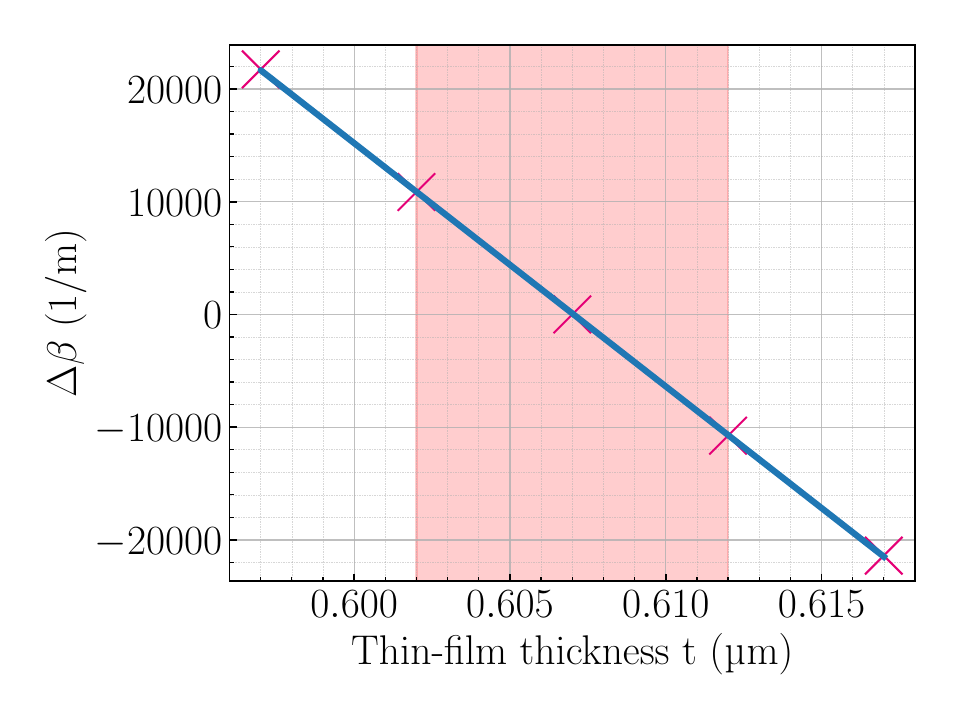}
        \label{fig:bild3}
    \end{subfigure}
    \begin{subfigure}{0.45\textwidth}
        \centering
        \includegraphics[width=\textwidth]{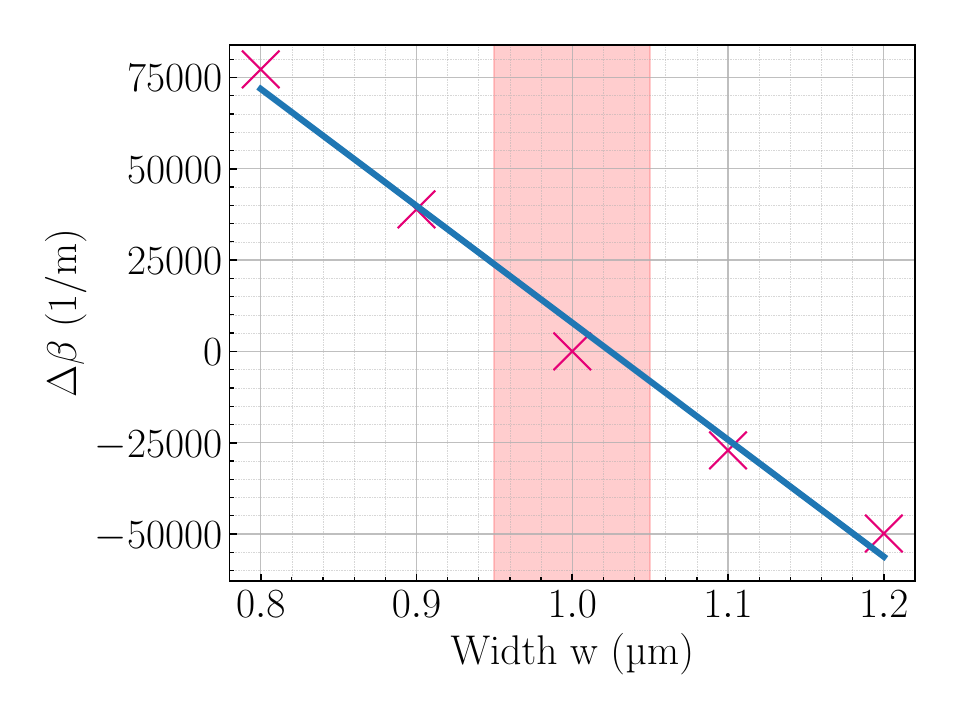}
        \label{fig:bild4}
    \end{subfigure}
    \caption{Influence of the individual geometry parameters on the phase mismatch such as angle, etching depth, thin-film thickness and width. A linear fit is applied to the data points obtained from Lumerical simulations. Furthermore, we mark in pink the tolerance regions of our fabrication.}
    \label{variation}
\end{figure}

\begin{figure}[h]
\centering\includegraphics[width=0.75\linewidth]{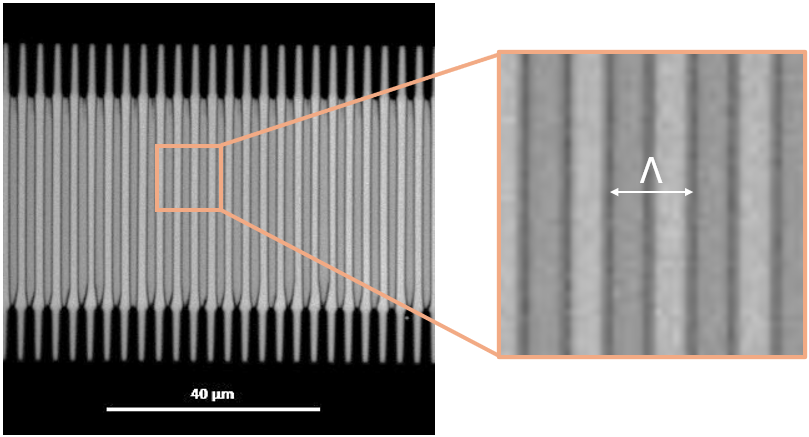}
\caption{Second-harmonic microscope image of a part of the 3$\,$mm long poled region. The individual poled domains can be identified. In the magnified view, a poling period is marked. }
\label{SEMimage}
\end{figure}

Our photon pair source is based on a type 0 parametric down-conversion process pumped with green light ($\sim$ 532$\,\mathrm{nm}$) that generates one photon in the visible ($\sim$ 810$\,\mathrm{nm}$), called signal, and the second in the telecom band ($\sim$ 1550$\,\mathrm{nm}$), called idler. Therefore, energy conservation is obeyed ($\hbar\omega_{p} = \hbar\omega_{s} + \hbar\omega_{i}$).
To achieve momentum conservation which is naturally not given for arbitrary wavelengths due to the material dispersion, we use quasi-phase matching obtained via periodic domain inversion. \cite{hum2007quasi}
The poling period $\Lambda$ is determined by the propagation constants:
\begin{align}
    \beta_{p} (\omega_{p}) - \beta_{s}(\omega_s) - \beta_{i}(\omega_i) - \frac{2 \pi}{\Lambda} = 0.
     \label{Polingperiod}
\end{align}
\noindent
To design our source we first need to find an appropriate waveguide geometry and suitable poling periods accordingly. We devise our waveguide geometry such that the waveguide only guides the fundamental mode for the longest wavelength and our poling periods take into account fabrication tolerances. Though higher order modes cannot be avoided for the shorter wavelengths, our design nevertheless minimizes the number of spatial modes for the generated photons.\\
In \autoref{modes} a) a schematic cross section of a TFLN waveguide is displayed. A waveguide is defined by the thin-film thickness $t$, the etching depth $h$, the top width $w$, the angle $a$ and the cladding thickness $c$.
The etching depth $h$ and angle $a$ are fixed by our fabrication process ($h=300\,\mathrm{nm}$ and $a=30\,$°). Moreover, we apply a SiO$_2$ cladding layer ($c=1000\,\mathrm{nm}$) to protect our waveguide from dust and other particles. For our 5$\%$ MgO-doped X-cut 1x1cm sample (NanoLN) we have measured the thin-film thickness with a nominal thickness of 600$\,\mathrm{nm}$ given by the manufacturer at five different points with an ellipsometer and average over these values to determine a mean thin-film thickness of our sample. The measured thin-film thickness is (607$\pm$1)$\,\mathrm{nm}$. This is crucial to know since the thin-film thickness has a significant influence on the poling period and thus on the generated output wavelengths \cite{kuo2022noncritical}. 
From all parameters only the waveguide width can be tuned to find a single mode waveguide at 1550$\,\mathrm{nm}$. For the given parameters we have simulated the TE modes at all wavelengths with the FDE solver of Lumerical and the Sellmeier equation according to the model reported in \cite{gayer2008temperature} and chose a waveguide top width of 1$\,\mathrm{\upmu m}$. In \autoref{modes} b)-d) the simulated fundamental TE modes for all three interacting fields are shown. 
\noindent
\\
For this geometry ($t=607\,\mathrm{nm}, h=300\,\mathrm{nm}, w=1\,\mathrm{\upmu m}, a=30\,$° and $c=1000\,\mathrm{nm}$) we calculate a poling period of $\Lambda=3.314\,\mathrm{\upmu m}$ using \autoref{Polingperiod}. 
The high confinement in TFLN rib waveguides leads to a modal dispersion and thereby propagation constants that are strongly dependent on the waveguide geometry. Differences between the fabricated and simulated geometries cause changes in the propagation constants that result in shifted wavelength combination for which \autoref{Polingperiod} is fulfilled. A minor variation of 1$\,\mathrm{nm}$ in the thin-film thickness can induce a $\sim$ 4$\,\mathrm{nm}$ shift in the telecom output wavelength.
To counteract for this, one could fabricate the waveguide, measure the geometry and match the poling period to the fabricated waveguide \cite{xin2025wavelength}. However, the LN thin-film is poled after etching resulting in a non-complete poling of the thin-film and thus a reduced efficiency. A different approach is to calculate a poling period range, as proposed in \cite{jankowski2021dispersion} to incorporate possible fabrication deviations.\\
Therefore, we evaluate the influence of each geometry parameter on the propagation constants individually to determine an appropriate poling period range based on our expected fabrication tolerances. \autoref{variation} illustrates the results of our simulations, which we used for finding an appropriate design.
\noindent
As an example, to estimate the influence of a deviating top width, we simulate the effective refractive index for different waveguide top widths ranging from 800-1200$\,\mathrm{nm}$ in steps of 100$\,\mathrm{nm}$. From these we calculate with \autoref{Polingperiod} the phase mismatch and use for the poling period $\Lambda = 3.31\,\mathrm{\upmu m}$ from the initial geometry. We plot the phase mismatch difference as a function of the top width and apply a linear fit to determine the slope $\frac{\delta \Delta \beta}{\delta w}$ (see \autoref{variation}). This analysis is performed analogously for all other parameters. The slopes of the individual parameters are as follows: 
\begin{align*}
    \frac{\delta \Delta\beta}{\delta a}=-2.3 \cdot 10^3 \frac{1}{\mathrm{m \, \text{°}}}, \frac{\delta \Delta\beta}{\delta h}=332.9 \cdot 10^3 \frac{1}{\mathrm{m \, \upmu m}},  \\  \frac{\delta \Delta\beta}{\delta t}=-2.2 \cdot 10^6 \frac{1}{\mathrm{m \, \upmu m}}, \frac{\delta \Delta\beta}{\delta w}=-320.2 \cdot 10^3 \frac{1}{\mathrm{m \, \upmu m}}.
\end{align*}
With these slopes we determine the phase mismatch:
\begin{align}
    \Delta \beta_{sum} &= \frac{\delta \Delta\beta}{\delta w} \cdot \Delta w + \frac{\delta \Delta\beta}{\delta h} \cdot \Delta h + \frac{\delta \Delta\beta}{\delta a} \cdot \Delta a + \frac{\delta\Delta \beta}{\delta t} \cdot \Delta t.
\end{align}
Here, $\Delta w$, $\Delta h$, $\Delta a$ and $\Delta t$ are the fabrication tolerances. Taking our fabrication tolerances of $\Delta w=100\,\mathrm{nm}$, $\Delta h=100\,\mathrm{nm}$, $\Delta a=5\,$° and $\Delta t=20\,\mathrm{nm}$ (marked as pink regions in \autoref{variation}), this leads to a phase mismatch of -54230$\,\frac{1}{m}$ and thus a wavelength shift of 100$\,\mathrm{nm}$ for the idler arm in the worst case. From these calculations, we can calculate a poling period range via the following equation:
\begin{align}
    \Delta \Lambda &= \frac{2 \cdot \pi}{\Delta \beta_0 \pm \Delta \beta_{sum}}.
\end{align}
\noindent
Here, $\Delta \beta_0$ is the phase mismatch of our initial geometry. According to the previous evaluation and the given parameters of our samples, we find that our design needs to consider a poling range between 3.266$\,\mathrm{\upmu m}$-3.443$\,\mathrm{\upmu m}$. Hence, we put the following five poling periods: 3.266$\,\mathrm{\upmu m}$, 3.310$\,\mathrm{\upmu m}$, 3.353$\,\mathrm{\upmu m}$, 3.397$\,\mathrm{\upmu m}$ and 3.443$\,\mathrm{\upmu m}$. With this poling period range we can be sure that we have at least one period that generates photons in the visible and telecom range. Moreover, we can also tune the generated output wavelength via the temperature. Based on our simulations, all manufacturing parameters are thus defined and we have found our design for suitable  periodically poled TFLN samples for our source. These are $t=607\,\mathrm{nm}, h=300\,\mathrm{nm}, w=1\,\mathrm{\upmu m}, a=30\,$° and $c=1000\,\mathrm{nm}$.\\
For the fabrication of our samples we start with the periodic poling of the thin film lithium niobate sample. To this aim, we first fabricate finger electrodes via a lift-off process using electron-beam lithography. For poling, we contact the 3$\,$mm long electrode and apply a single high-voltage pulse to invert the spontaneous polarisation in the crystal. In \autoref{SEMimage} a second-harmonic microscope image of a part of the poled region is shown. Following the poling process, waveguides are fabricated within the poled region using a combination of electron-beam lithography and physical dry etching. Finally, a 1$\,\mathrm{\upmu m}$ SiO$_2$ cladding layer is deposited via plasma-enhanced chemical vapor deposition.
In the final step, the end facets of the sample are chemomechanically polished to improve the free-space coupling efficiency into the waveguides. 
The final waveguide has a length of 7$\,$mm.\\
We measure the losses of our waveguides via the Fabry-Pérot method \cite{regener1985loss}. For this method we use the reflectivity of bulk LN. The losses of our poled waveguides are around 1$\,$dB/cm.

\section{Experimental Setup}
\label{Setuptext}

\begin{figure}[htbp]
\centering\includegraphics[width=\linewidth]{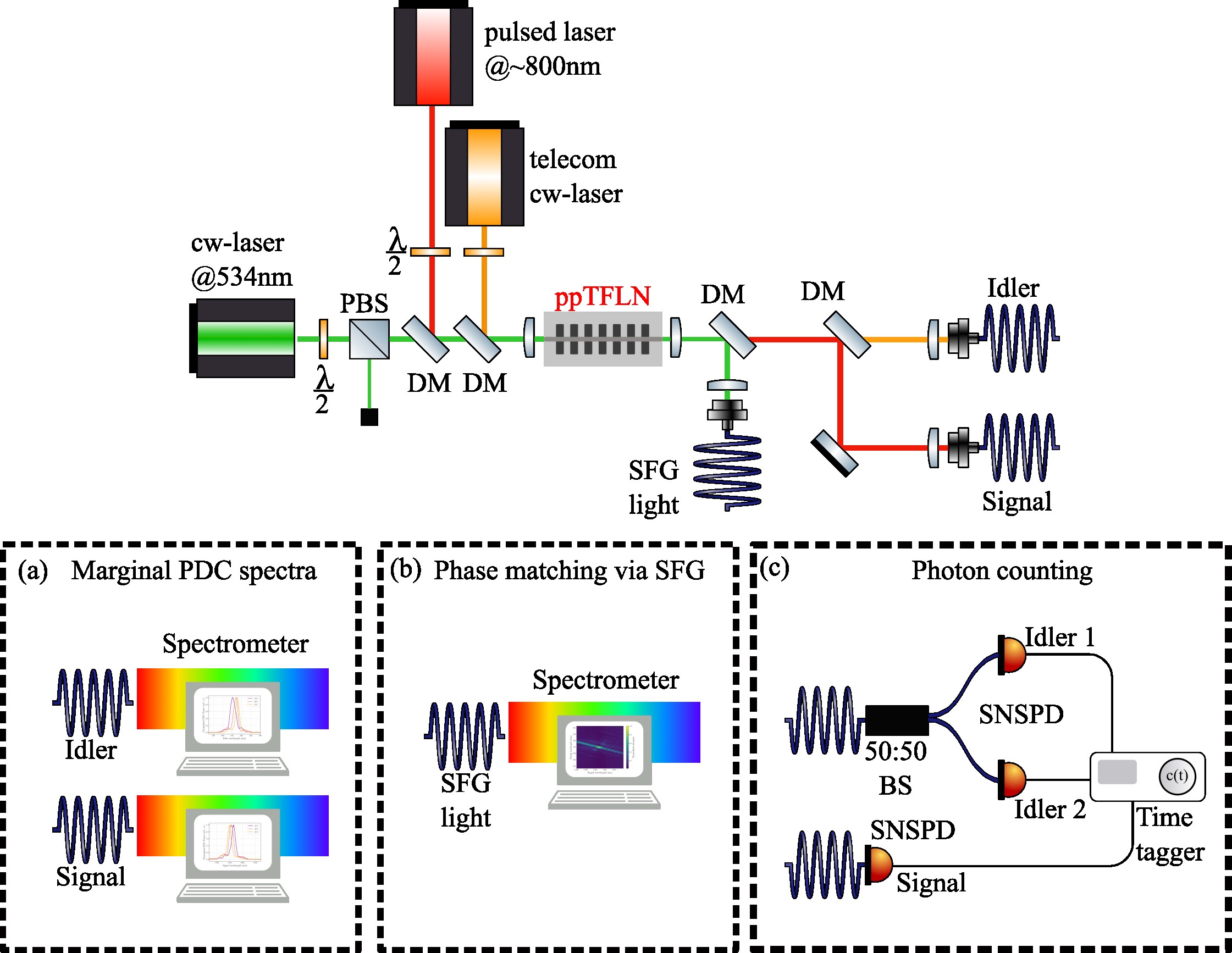}
\caption{Experimental setup for source characterisation. The setup consists of the source part and three different measurement setups. a) Setup to measure the spectral characteristics of the PDC process. b) Setup to measure the phase matching of the waveguide by using a sum-frequency generation process. c) Hanbury-Brown Twiss setup for analysing the photon number purity and related properties. $\frac{\lambda}{2}$ : half wave plate, PBS: polarizing beam splitter, DM: dichroic mirror.}
\label{Setup}
\end{figure}

Our optical experimental setup for the implementation and characterization of the two photon pair source is shown in \autoref{Setup} . It consists of the source part with the generation of the photon pairs followed by three different measurement configurations for the evaluation of the performance characteristics. Two of them examine the spectral characteristics of the photon pairs and the third one analyses the quality of the photon number statistics and the purity of heralded single photon states in dependence of the used pump powers.\\
The source is pumped by a tunable cw-laser at 534$\,$nm. To control the polarization and the power for our PDC process we use a half wave plate ($\lambda / 2$) and subsequent a polarising beam splitter (PBS). 
\noindent For analysing the phasematching we employ the reverse sum frequency process and launch a pulsed Ti:sapphire laser (795$\,\mathrm{nm}$ - 825$\,\mathrm{nm}$) in the visible and a telecom cw-laser (1500$\,\mathrm{nm}$-1600$\,\mathrm{nm}$). Both are incorporated into the main beam path via dichroic mirrors.
For coupling into and out of our periodically poled thin-film lithium niobate (ppTFLN) waveguide we use aspheric lenses. Due to the coupling with single lenses the coupling is aperture limited and thus determining a coupling efficiency is challenging.
After the ppTFLN waveguide, we employ two dichroic mirrors, where one is used as a low-pass filter to separate the generated photon pairs from the pump light and a second dichroic mirror separates the signal photon from the idler photon.
All three wavelengths are finally coupled into three different single mode fibers.\\
For measurements of the spectral characteristic we use a single photon sensitive spectrometer consisting of a spectrograph combined with a CCD camera to measure the signal spectrum and another single photon sensitive spectrometer consisting of a spectrograph with a iDus InGaAs camera for the idler arm (\autoref{Setup} a)).
Moreover, when investigating the phase matching via SFG we use the pulsed Ti:sapphire laser and the telecom cw-laser and measure the generated light with the same single photon sensitive spectrometer as used for measuring the signal spectrum connected to the fiber of the green light (\autoref{Setup} b)).\\
For our photon counting measurements we study and analyse the coincidence counts in dependence of the pump power and realize a Hanbury-Brown Twiss setup for obtaining more detailed information of higher photon number contributions (\autoref{Setup} c)). Thus, we introduce a fiber 50:50 beam splitter into the idler arm. Then, we connect the two outputs of the fiber beam splitter called idler1 and idler2 to superconducting nanowire single photon detectors (SNSPDs) as well as the signal output. The efficiencies of the detectors are 83$\,\%$, 83$\,\%$ and 85$\,\%$, respectively. We analyse the data from the SNSPDs with a time tagger and measure the single clicks of all three channels as well as the coincidence between idler1 and signal and idler2 and signal and the threefold coincidence between all three channels. For these measurements we use a coincidence window of 1$\,$ns.

\section{Source Characterisation}
\label{SourceCharac}
We start the characterisation of our source with analysing the spectral properties of generated photon pairs. Moreover, we analyse the quantum performance by analysing the photon counting experiments.

\subsection{Spectral Characteristics}

\begin{figure}[h]
\centering\includegraphics[width=7cm]{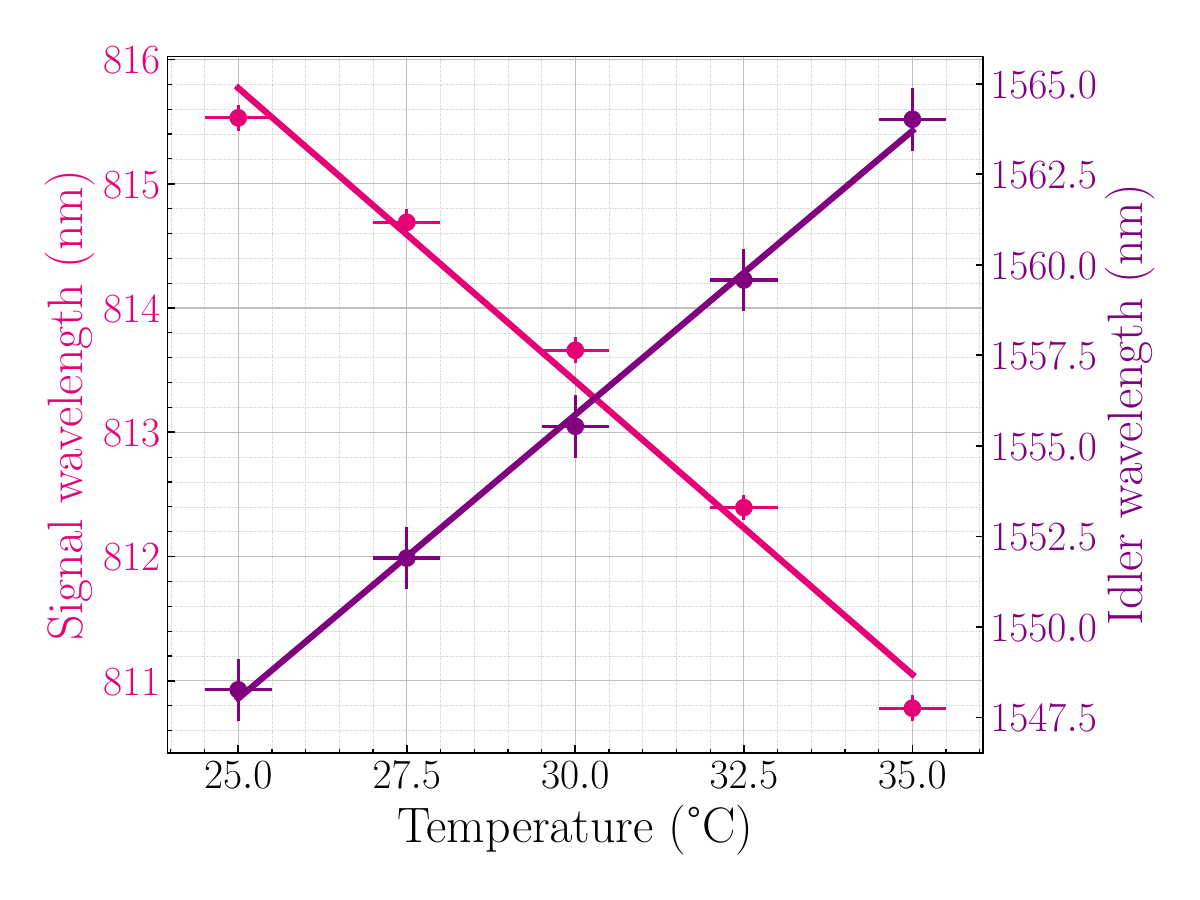}
\caption{Temperature dependency of the peak wavelength for the signal and idler peak wavelengths. With increasing temperature, the idler wavelength increases and the signal wavelength decreases.}
\label{Temperature}
\end{figure}

With the setup shown in \autoref{Setup} a), the dependency of the generated output wavelength on the poling period is measured. The source is pumped with a wavelength of 534$\, \mathrm{nm}$. We change the pump wavelength slightly to a value of 534$\, \mathrm{nm}$ as the process shifts for a different real waveguide geometry as compared to the one assumed in the simulations. By changing the pump wavelength the generated output wavelength shift for a fixed poling period leading to a generation of the photons at our desired wavelengths.
\begin{figure}[htbp]
\centering\includegraphics[width=\linewidth]{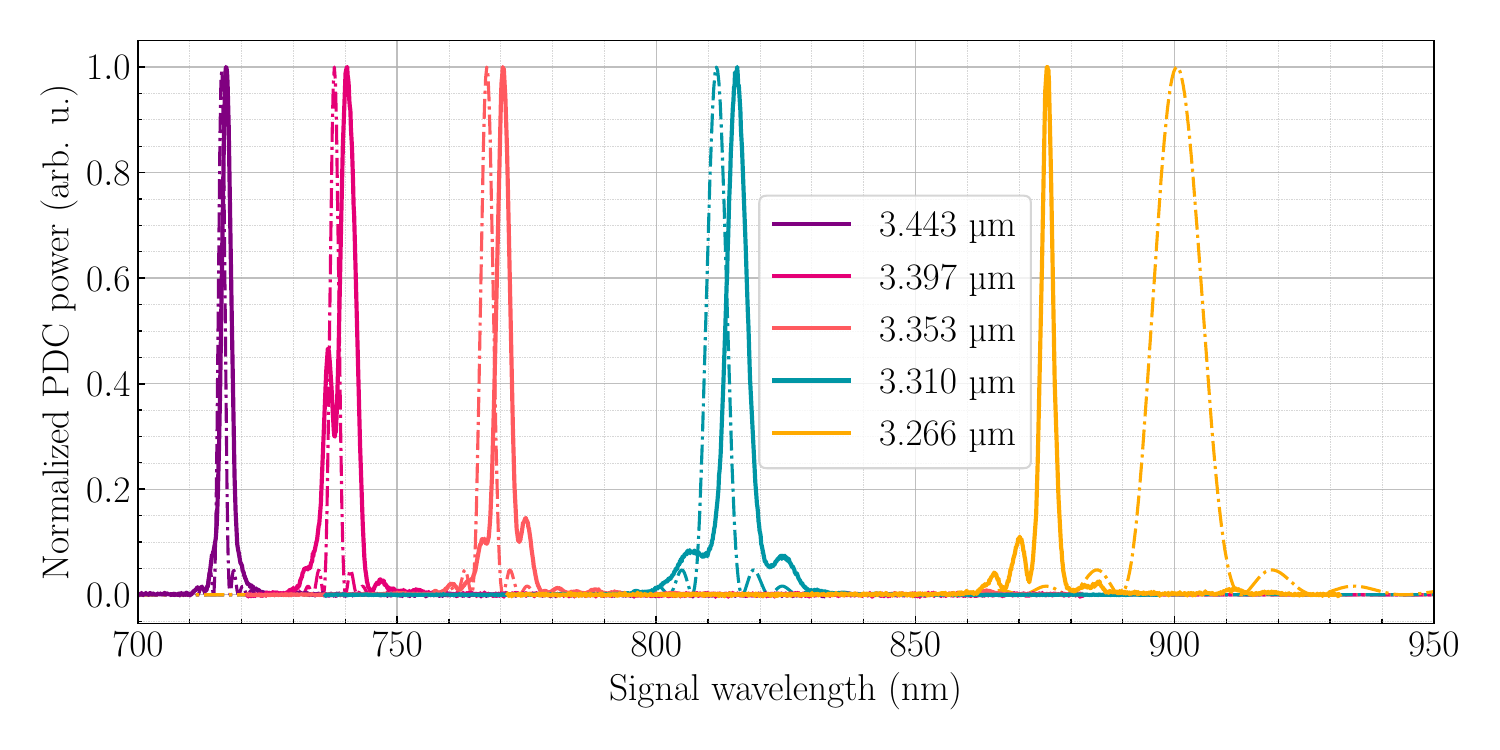}
\caption{Normalized PDC spectrum of different waveguides with different poling periods ranging from 3.266$\,\mathrm{\upmu m}$ to 3.443$\,\mathrm{\upmu m}$ at a temperature of 25$\,$°$\mathrm{C}$. The solid line represent the measurement and the dashed line the simulation.}
\label{poling}
\end{figure}
In \autoref{poling} the normalized PDC spectra of five different waveguides with the same geometry but different poling periods as a function of the signal wavelength are shown. Decreasing the poling period leads to an increased signal wavelength, in strong agreement with our simulations (dotted line). The experimental data exhibits a slight shift in wavelengths and bandwidth especially the waveguide with a poling period of 3.266$\,\mathrm{\upmu m}$ that can be attributed to deviations in the device geometry as well as alignment and chromatic aberrations.
The waveguide with a 3.266$\,\mathrm{\upmu m}$ poling period (yellow) exhibits more pronounced side peaks on the left, while for the 3.31$\,\mathrm{\upmu m}$ waveguide (turquoise) peaks on both sides can be seen. The 3.443$\,\mathrm{\upmu m}$ waveguide (magenta) has a narrower bandwidth than the 3.353$\,\mathrm{\upmu m}$ waveguide (orange). We assume that this behaviour is not due to a different poling quality but rather that this is an effect of a varying thin-film thickness along the waveguide. These variations are not statistically distributed, but are rather long range noise which could lead to the above discussed effects \cite{santandrea2019general}.\\
For our further experiments we choose the waveguide with a poling period of 3.31$\,\mathrm{\upmu m}$ since this generates photons at the desired wavelengths of 815$\,\mathrm{nm}$ and 1550$\,\mathrm{nm}$. 
\begin{figure}[htbp]
\centering\includegraphics[width=8cm]{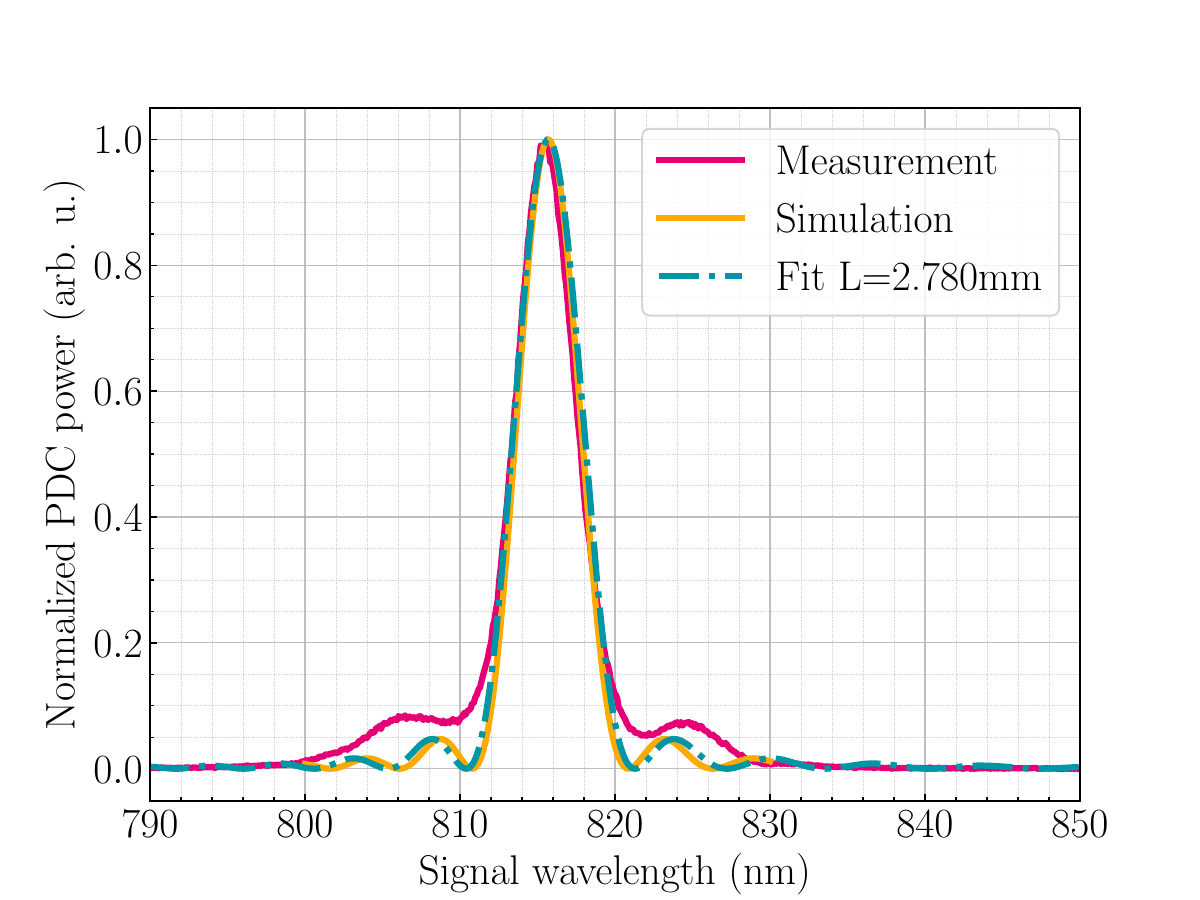}
\caption{Idler spectrum measured with a single-photon sensitive spectrometer of a waveguide with a poling period of 3.31$\,\mathrm{\upmu m}$ at a temperature of 25$\,$°C in magenta. Furthermore, we show a simulated spectrum of a 3$\,$mm poled waveguide in yellow and a sinc-fit to determine the effective poling length in turquoise.}
\label{Effectivelength}
\end{figure}
\noindent
Thus, for this waveguide we analyse the bandwidth and shape in more detail (see \autoref{poling}). The bandwidth of a PDC spectrum depends on the length of the periodically poled waveguide. A longer poling region leads to a narrower line width.
However, this strongly depends on the fabrication tolerances and the so-called effective poling length indicates the quality of poling and waveguide homogeneity. To quantify the broadening of our spectrum, we determine the effective poling length. Therefore, we plot in \autoref{poling} the signal spectrum of the waveguide with a poling period of 3.31$\,\mathrm{\upmu m}$ (see \autoref{poling}, magenta) and fit a sinc-function to the measurement (turquoise). In yellow, we plotted the simulated spectrum. Comparing simulation and measurement, our measurement resembles a near to perfect sinc-shape. An effective poling length of 2.78$\,\mathrm{mm}$ is determined. The effective poling length of 2.78$\,\mathrm{mm}$, corresponding to 93$\,\%$ of the nominal design, highlights the high fabrication quality and precise domain engineering achieved in our device. Such a high effective poling length, close to the nominal design, indicates good fabrication accuracy and suggests that the source performance is not significantly limited by poling imperfections.
We attribute the minor broadening of the spectrum to long range noise in the thin-film thickness along the waveguide leading to changes in the dispersion and thus to a broadening of the PDC spectrum since the phase matching condition is fulfilled for different wavelength combinations  \cite{chen2024adapted}. 
The highest impact on the dispersion can be attributed to the thin-film thickness \cite{kuo2022noncritical} which can vary over the sample. To compensate for the thin-film thickness variation one can either use adaptive poling \cite{chen2024adapted} or adapt the waveguide width \cite{he2024efficient}.\\
As a next step, we characterise the shift of the generated output wavelengths for different temperatures of this waveguide. 
The peak wavelength of signal and idler for different temperatures from 25$\,$°C to 35$\,$°C in 2.5$\,$°C steps are shown in \autoref{Temperature}. The signal wavelength decreases with increasing temperature and the idler wavelength increases with increasing temperature. We measure a temperature dependency of -0.472$\,$nm/K for the signal and 1.57$\,$nm/K for the idler photons. Simulations lead to values of -0.61$\,$nm/K for the signal and 2.28$\,$nm/K for the idler arm showing that the general trend is captured, but the values themselves deviate. The difference can be attributed to minor inaccuracies in the simulated dispersion as well as a temperature difference between the set temperature of the heater and the actual temperature of the sample.  With this temperature tunability our source is adaptable to different applications with specific wavelength.
\noindent
\\
Lastly, we map the 2D phase matching function of the source. To this end we construct a contour plot for measured counts in dependence of signal and idler wavelengths.
\begin{figure}[h]
\begin{subfigure}[h]{0.49\linewidth}
\includegraphics[width=\linewidth]{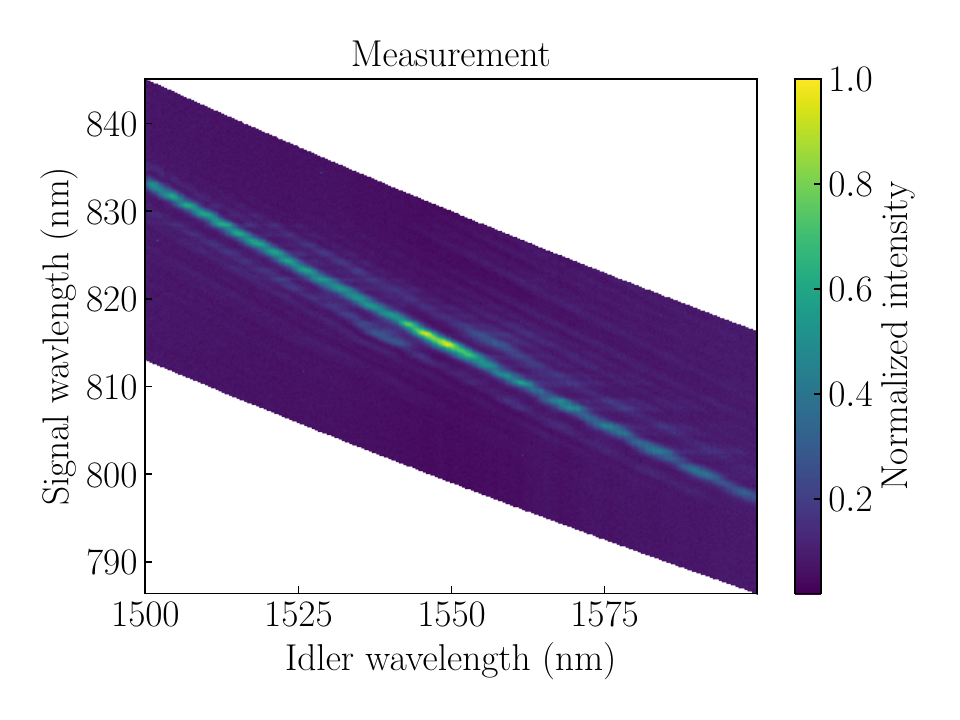}
\end{subfigure}
\hfill
\begin{subfigure}[h]{0.49\linewidth}
\includegraphics[width=\linewidth]{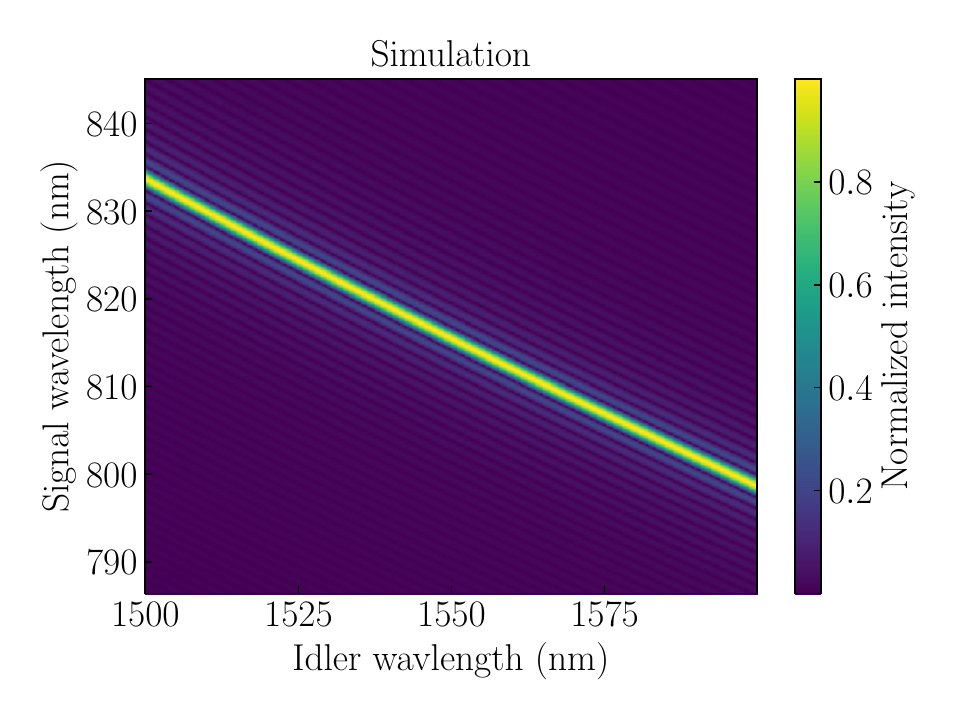}
\end{subfigure}%
\caption{Measured and simulated phase matching spectrum of the PDC process. The measurement is in good agreement with our simulation.}
\label{phase matchingscan}
\end{figure}
\noindent
We utilize a sum-frequency generation process and scan the pulsed Ti:sapphire laser from $795\,$nm to $825\,$nm in steps of $5\,$nm. For each wavelength, we scan the tunable cw-laser from $1500\,$nm to $1600\,$nm in steps of  $0.2\,$nm and record the spectrum of the generated SFG light on the single photon sensitive spectrometer. Finally, we combine these measurements to get the phase matching function.
The left plot in \autoref{phase matchingscan} shows the measurement. We see a brighter spot at 1550$\,$nm and 815$\,$nm since the setup was aligned for this wavelength combination. We measured at this point around 3500 counts with an integration time of 1$\,$s. The dark counts are around 50 counts. The right plot shows the simulation and when comparing our measurement to the simulation we find a very good agreement with the experiment in general. Note that this confirms nicely that our simulations are highly suitable for the design of adaptable and customized photon pair sources according to desired spectral properties. Again, a slight offset between measurement and simulation can be seen, but the simulated phase matching angle of -51.66$\,$° and the measured one of -52.60$\pm$0.11$\,$° match perfectly.
We attribute the difference to the high sensitivity of the dispersion on the waveguide geometry.

\subsection{Photon Number Purity}
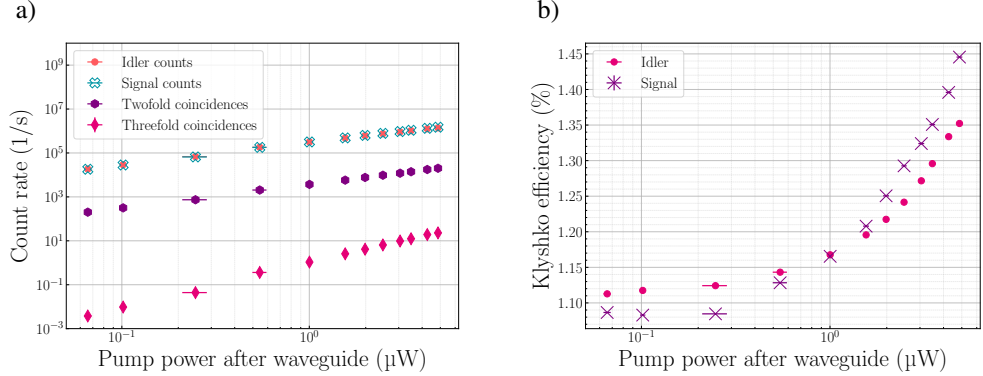
\begin{figure}[h]
\centering
\begin{subfigure}[h]{0.48\linewidth}
    \begin{picture}(0,0)
        \put(10,140){\text{a)}} 
    \end{picture}
    \includegraphics[width=\linewidth]{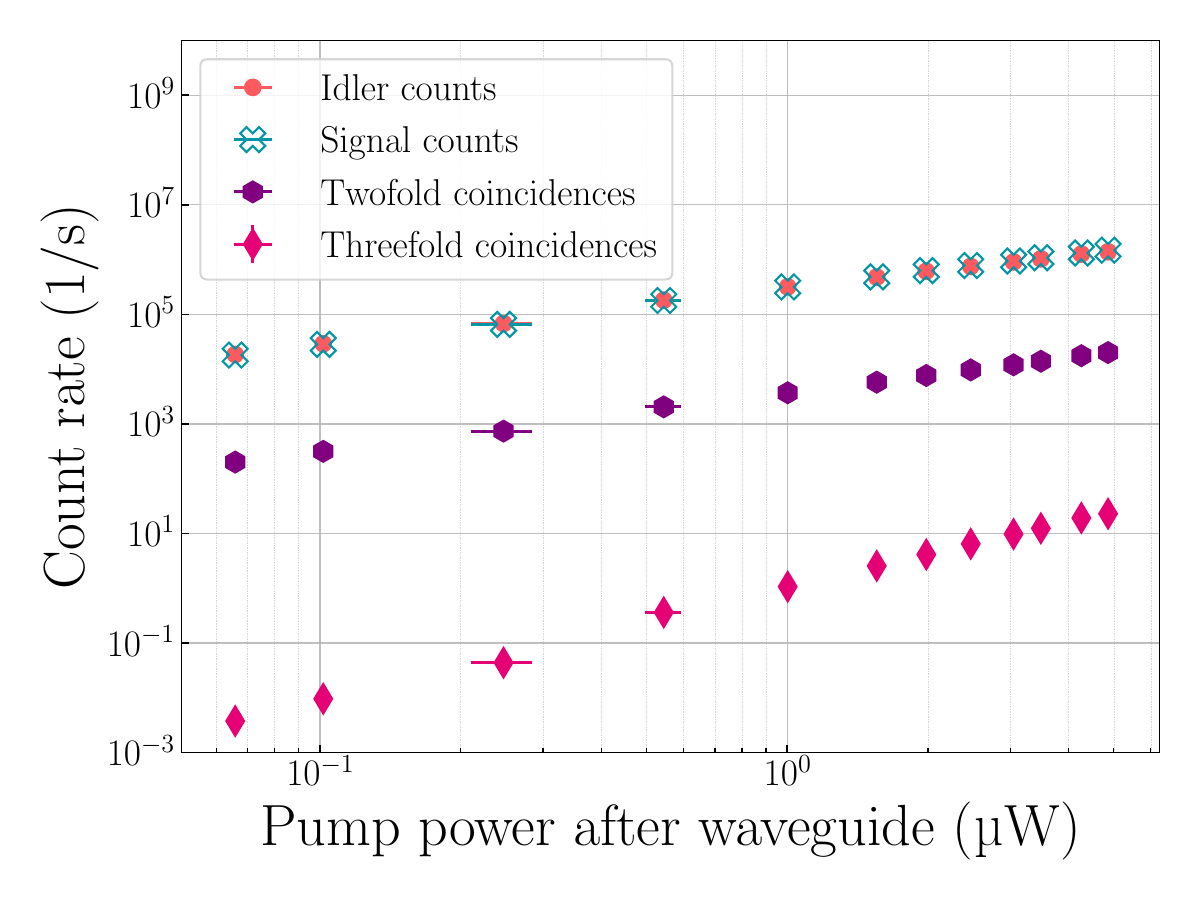}
\end{subfigure}
\hfill
\begin{subfigure}[h]{0.48\linewidth}
    \begin{picture}(0,0)
        \put(10,140){\text{b)}} 
    \end{picture}
    \includegraphics[width=\linewidth]{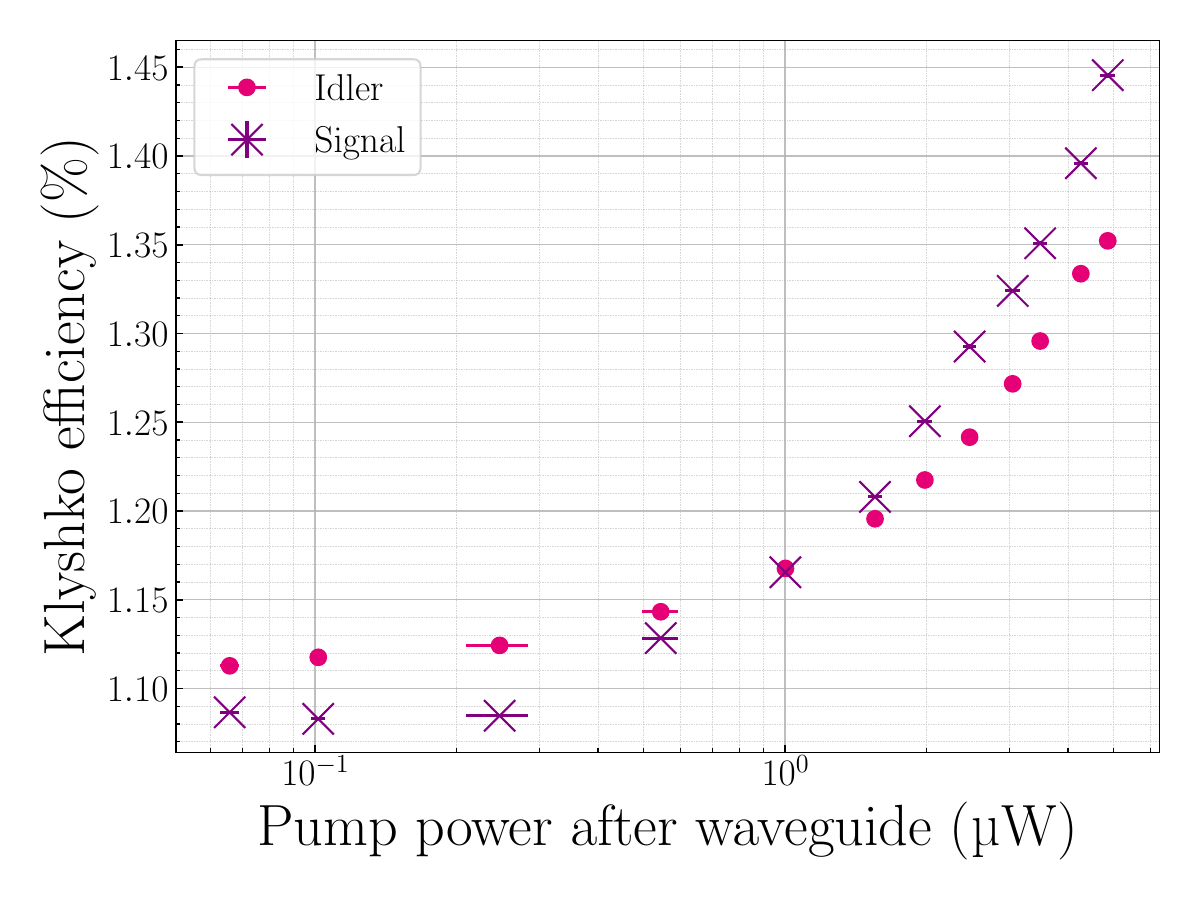}
\end{subfigure}
\caption{a) Pump power dependent count rates. The idler and signal counts as well as the twofold coincidences increase linearly with the power of 0.92, 0.96 and 1.08, respectively. The threefold coincidences increase superlinearly with the power of 1.85. b) Klyshko efficiency for signal and idler as a function of the pump power measured after the waveguide.}
\label{countsmeanphoton}
\end{figure}

\begin{figure}[h]
\centering
\begin{subfigure}[h]{0.48\linewidth}
    \begin{picture}(0,0)
        \put(10,140){\text{a)}} 
    \end{picture}
    \includegraphics[width=\linewidth]{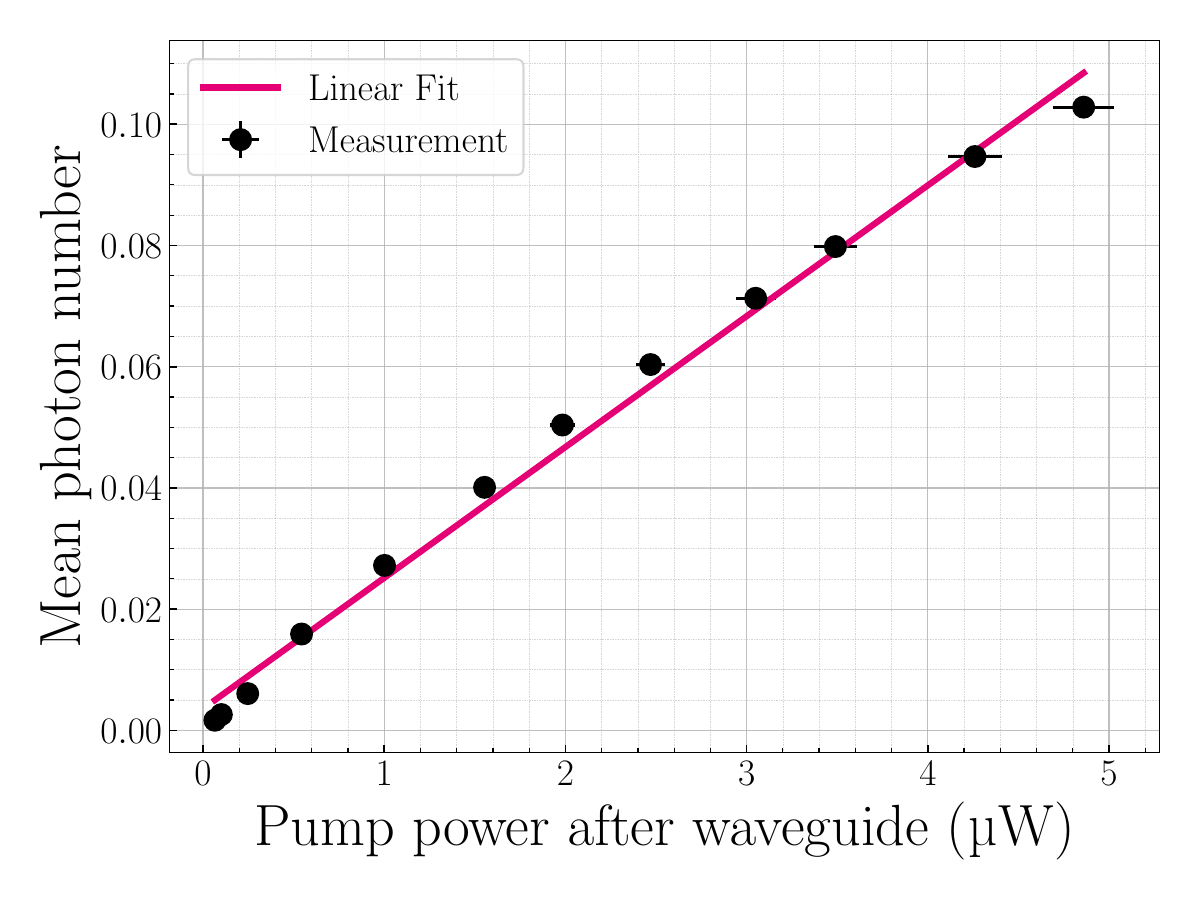}
\end{subfigure}
\hfill
\begin{subfigure}[h]{0.48\linewidth}
    \begin{picture}(0,0)
        \put(10,140){\text{b)}} 
    \end{picture}
    \includegraphics[width=\linewidth]{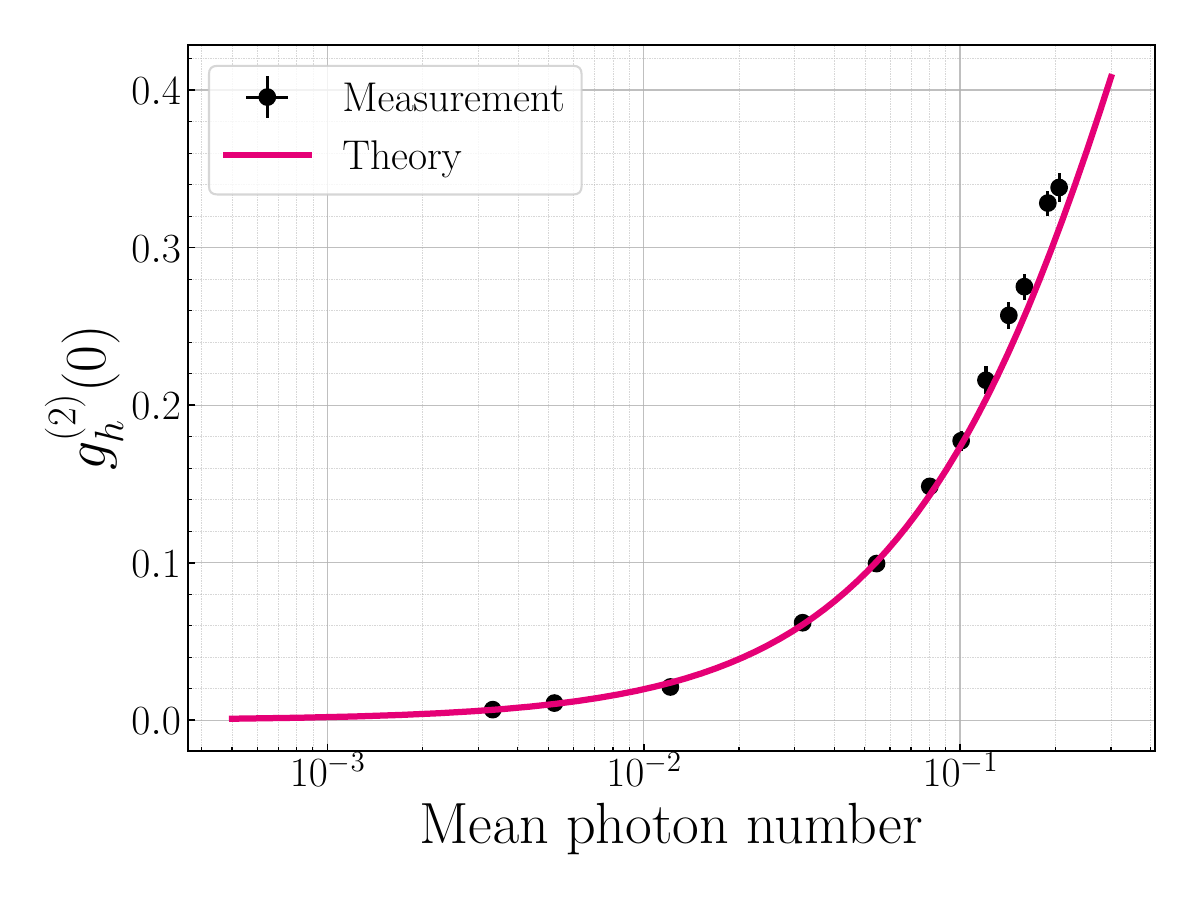}
\end{subfigure}
\caption{a) Mean photon number as a function of the pump power.  b) Heralded second-order correlation function as function of the mean photon number.}
\label{g2andklyhsko}
\end{figure}
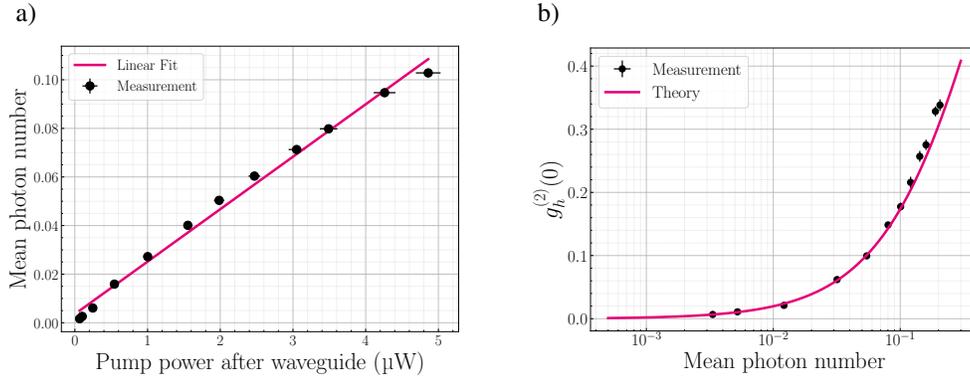

For studying the performance of our source in terms of quantum photon statistics we perform different photon counting experiments.
First, we measure the single count rates of signal and idler and the threefold coincidences between signal, idler1 and idler2 for different pump powers with the setup as depicted in \autoref{Setup} (c). 
The pump power is measured with a powermeter in front of the fiber coupling of the pump light after the waveguide.
The pump power dependent counts are plotted in \autoref{countsmeanphoton} a).
\noindent The measured two  fold coincidences show that we are generating correlated photon pairs. 
From these count rates, we first calculate the Klyshko efficiencies \cite{klyshko1980use}. The Klyshko efficiency can be used as a measure of the detection efficiency of our generated photons from the source to our detector. It describes the amount of pairs detected compared to single photons and it is reduced by single photon loss. Moreover, this is an important quantity since by determining the Klyshko efficiency, we can calculate the actual generated pairs inside the source from the measured counts. We calculate this via the following relation:
\begin{align}
    \eta_{signal} = \frac{C_{s,1} + C_{s,2}}{N_{1} + N_{2}} \\
    \eta_{idler} = \frac{C_{s,1} + C_{s,2}}{N_{s}}.
\end{align}
\noindent
Here, $C_{s,1}$ describes the coincidences between signal and idler1 and $C_{s,2}$ between signal and idler2. $N_{1}$ are the single idler1 counts, $N_{2}$ the single idler2 counts and $N_{s}$ are the single signal counts. 
In \autoref{countsmeanphoton} b) the Klyhsko efficiencies are plotted as a function of the pump power. 
With increasing pump power the efficiencies slightly increase due to higher-order photon number components. The increase for the signal and idler photons is different which can be attributed to different losses inside the waveguide for the signal and idler wavelength.  For low pump powers the Klyshko efficiency is almost constant with 1.1$\,\%$ for the idler photon and 1.08$\,\%$ for the signal.
These efficiencies may seem rather low, however they include all setup and detection losses and are obtained with lens coupling to the waveguide. 
By improving the in- and outcoupling of the waveguide by for example waveguide tapers or grating couplers, the efficiencies can be improved significantly \cite{hu2021high,huang2024low}.\\
The increase of the Klyshko efficiency as well as the increase of the two and threefold coincidences shows that the source generates multiple photon pairs. To investigate this further, we determine the second-order correlation function $g^{(2)}_h(0)$ since this is a figure of merit for the simultaneous generation of multiple photon pairs. Based on this definition, the following relation can be derived to describe this behaviour theoretically \cite{laiho2011characterization}:
\begin{align}
    g^{(2)}_h (0,n) = \frac{\sum_{\Tilde{n}} \Tilde{n}^2 \left(\Tilde{n} - 1 \right) \rho\left(\Tilde{n}\right)_{PDC}}{\left[ \sum_{\Tilde{n}} \Tilde{n}^2 \rho\left(\Tilde{n}\right)_{PDC} \right]^2}
\end{align}
\noindent
where $n$ is the mean photon number per coincidence window and $\Tilde{n}$ is the photon number running from zero to infinity. 
For $\rho\left(\Tilde{n}\right)_{PDC}$ we assume Poissonian statistics \cite{avenhaus2008photon},
\begin{align}
    \rho \left( \Tilde{n} \right) _{PDC} = \frac{n^{\Tilde{n}}}{\Tilde{n}!} \cdot e^{-n}
\end{align}
\noindent
because of the multitude of spectral modes in our source. To compare our measurement to the theory, we first calculate the mean photon number. For this, we multiply the PDC rate by the coincidence window of 1$\,$ns that we set for the time tagger. In \autoref{g2andklyhsko} a) the mean photon number as a function of the pump power is shown and we apply a linear fit. It can be seen that for high pump power, the mean photon number starts to deviate from the linear behaviour which we attribute to detector saturation effects. We reach mean photon numbers between 1.66$\cdot$10$^{-3}$ and 103$\cdot10^{-3}$.\\
Next, we calculate the $g^{(2)}_h(0)$ via the following equation:
\begin{align}
    g^{(2)}_h(0) = \frac{{C_{s,1,2}} \cdot N_{s}}{C_{s,1} \cdot C_{s,2}}.
\end{align}
\noindent
Here, $C_{s,1,2}$ denotes the threefold coincidences, $N_{s}$ the single signal counts and $C_{s,1}$ / $C_{s,2}$ the twofold coincidences between signal and idler 1 or idler 2.  
For a perfect heralded single photon pair source this value should be zero, since no threefold coincidence should be measured if only one photon pair is generated. 
By increasing the pump power the probability of generating multiple photon pairs increases and thus the $g^{(2)}_h(0)$ value increases.
\noindent
In \autoref{g2andklyhsko} b) the $g^{(2)}_h(0)$ values as a function of the mean photon number are shown.
For low mean photon numbers we obtain a $g^{(2)}_h(0)$ value of $(6.7\pm1.1)\cdot$10$^{-3}$.
This confirms that our source is able to generate high quality heralded single photons. We can also see that theory and experiment fit perfectly.\\
As a last step we investigate the brightness of our source.

\subsection{Brightness}

The brightness $B$ of a PDC source pumped with cw light is typically defined as the generated photon pair rate $R_{PDC}$ over the pump power inside the waveguide $P$ and the bandwidth $\Delta \nu$ of the generated photons:
\begin{align}
    B = \frac{R_{PDC}}{P\cdot \Delta \nu}.
\end{align}
\noindent
This corresponds to the spectral brightness of heralded single photons for strongly correlated photon pairs.
Since we only measure the pump power after the waveguide we calculate the power pump inside the waveguide by estimating the out-coupling efficiency and take the reflection from the end-facet and the lens into account. For the end-facet reflectivity we take the standard value of LN of 14.54$\,\%$ at 532$\,\mathrm{nm}$ and for the lens we use a reflectivity of 5$\,\%$. For the coupling we estimate by simulating the overlap between the waveguide mode and the mode distribution of the lens of around 40$\,\%$. This means we have a overall transmission from waveguide to the power meter of 32$\,\%$ which we use as correction factor. From this, we calculate the actual pump power inside the waveguide. For a pump power after the waveguide of $P_{\text{power}}$=544$\,\mathrm{nW}$, we therefore arrive at a maximal pump power in the waveguide of $P_{\text{in}}$=1.68$\,\mathrm{\upmu W}$. The PDC rate adjusted with the Klyshko efficiency for this pump power is $R_{PDC}$=15.89$\,\mathrm{MHz}$. The bandwidth of our generated photons is around $\Delta \omega$=2104$\,\mathrm{GHz}$.
This yields a brightness of 0.44$\cdot10^7\,\frac{\text{pairs}}{\text{s} \cdot \text{mW} \cdot \text{GHz}}$.\\
We can now compare this brightness to other sources published in literature (see \autoref{tabelle}).

\definecolor{headerColor}{HTML}{0096A5} 
\definecolor{rowColor}{HTML}{D6E3E3}    
\definecolor{lastRowColor}{HTML}{ffd47f}
\definecolor{RowColor}{HTML}{d3d3d3}

\begin{table}[h]
    \caption{Comparison of brightness of different PDC sources in lithium niobate.}
    \centering
    \renewcommand{\arraystretch}{1.3}
    \setlength{\arrayrulewidth}{1pt} 
    \arrayrulecolor{headerColor} 

    \begin{tabular}{|p{7cm}|c|c|}

        \hline
        \rowcolor{headerColor} 
        \textbf{\textcolor{white}{Type of Source}} & \textbf{\textcolor{white}{Brightness $\left[ \frac{\text{pairs}}{\text{s} \cdot \text{mW} \cdot \text{GHz}} \right]$}} & \textbf{\textcolor{white}{\textbf{Citation}}} \\
        \hline
         Bulk LN source @1064~nm& 2.5$\cdot$10$^4$  & \cite{tabakaev2022spatial} \\
        \hline
        Dispersion engineered LN waveguide source @1350~nm& 1.5$\cdot$10$^5$ & \cite{pollmann2024integrated} \\
        \hline
        Single pass TFLN waveguide soure @1475~nm& 1.1$\cdot$10$^7$ & \cite{xue2021ultrabright} \\
        \hline
        \rowcolor{RowColor}
        Resonant TFLN waveguide source @1550~nm& 1.6$\cdot$10$^{10}$ &\cite{ma2023highly} \\
        \hline
        \rowcolor{lastRowColor}
         This work (single pass waveguide source)& 0.44$\cdot$10$^{7}$ & \\
        \hline
    \end{tabular}
    \label{tabelle}
\end{table}
\noindent
Comparing the brightness of our source to a Type-0 bulk LN source pumped at 532$\,\mathrm{nm}$ to generate photons at 1064$\,\mathrm{nm}$, we achieve an increase in the brightness of 3 orders of magnitude. This increase is due to the fact that our source features waveguides. The use of waveguides leads to an increase in the field overlap of the interacting fields and a longer interaction length which both increase the brightness.\\
Furthermore, the brightness of our source is two orders of magnitudes brighter compared to a dispersion engineered LN waveguide source generating photons at 1350$\,\mathrm{nm}$ \cite{pollmann2024integrated} since the titanium in-diffused waveguides provide a lower index contrast compared to the TFLN waveguides. Due to the higher index contrast and the reduced size of the TFLN waveguides leading to higher confinement, the interacting fields have an even higher overlap which increases the brightness.\\
Hence, we compare our source to already published PDC sources in TFLN. Here, we can see that our single pass TFLN waveguide source has a comparable brightness to other publications \cite{xue2021ultrabright}. However, the brightness strongly depends on the pump power inside the waveguide which is challenging to measure. To further increase the brightness resonant schemes instead of a single pass configuration could be utilised as reported in \cite{ma2023highly} with a brightness of 10$^{10}$~$\frac{\text{pairs}}{\text{s} \cdot \text{mW} \cdot \text{GHz}}$. Nonetheless, compared to single pass sources in TFLN our source yields a brightness in the same order of magnitude while generating photons at two distinct wavelengths.

\section{Conclusion}
\label{Conclusion}

To conclude, we demonstrate an integrated highly frequency non-degenerate photon pair source in TFLN. With its high spectral brightness and single photon quality this source is a fundamental building block for various integrated quantum optics experiments in the fields of quantum communication or quantum spectroscopy.\\
The output wavelengths of the source are at $\lambda_i=1550\,\mathrm{nm}$ and $\lambda_s=815\,\mathrm{nm}$ and can be tuned via temperature with 1.57$\,\mathrm{nm/K}$ for the idler photon and -0.472$\,\mathrm{nm/K}$ for the signal photon. This tunability makes the source adaptable to different applications with specific wavelength requirements. We observe a sinc-like phase matching curve which indicates an excellent poling homogeneity and uniformity of the device over the whole length. The close match between the effective and nominal poling length underscores the consistent approach taken from device design through fabrication to quantum characterization, enabling the optimization of high-quality quantum state generation. Additionally, we measure the 2D phase matching and conclude that we can accurately simulate the dispersion of our source.\\
Furthermore, we analysed the generated photon pairs by performing photon counting experiments at different pump powers. We determine a heralded second-order correlation function of $g^{(2)}_h(0) = (6.7\pm1.1)\cdot10^{-3}$. 
This allows to extract high quality heralded single photons.
Moreover, we calculate a high spectral brightness of 0.44$\cdot10^7$~$\frac{\text{pairs}}{\text{s} \cdot \text{mW} \cdot \text{GHz}}$ which is comparable to other single pass sources in TFLN. We compare simulations and measurements of the spectral and quantum characteristics, demonstrating precise control over the PDC process in our device.\\
To conclude, we successfully developed a photon pair source in TFLN generating two photons bridging the gap between the visible and infrared spectral regions and enabling new possibilities for hybrid quantum systems. Due to the enhanced nonlinear interaction of the nanostructured waveguides our source exhibits a high brightness. This source is an ideal candidate for various applications in the field of quantum optics and a fundamental building block for QPICs.\\

\begin{backmatter}
\bmsection{Funding}
Deutsche Forschungsgemeinschaft (DFG, German Research Foundation) – SFB-Geschäftszeichen TRR142/3-2022 – Projektnummer 231447078; Federal Ministry of Education and Research / Bundesministerium für Bildung und Forschung (Grant number 13N15975); Max Planck School of Photonics.

\bmsection{Acknowledgments}
Silia Babel and Franz Roeder are part of the Max Planck School of Photonics supported by the German Federal Ministry of Education and Research (BMBF), the Max Planck Society, and the Fraunhofer Society.

\bmsection{Disclosures}
The authors declare no conflicts of interest.

\bmsection{Data Availability Statement}
Data underlying the results presented in this paper are not publicly available at this time but may be obtained from the authors upon reasonable request.

\end{backmatter}


\bibliography{sample}

\end{document}